# Exploring Scavenging Strategies and Cognitive Problem-Solving in Indian Free-Ranging Dogs

Tuhin Subhra Pal[1], Srijaya Nandi[1†], Hindoli Gope[1†], Aniket Malakar[2], Rohan Sarkar[1,] Sagarika Biswas[1], Anindita Bhadra[1*]

[1] Department of Biological Sciences, Indian Institute of Science Education and Research, Kolkata, Mohanpur, West Bengal 741246, India

[2] Department of Biotechnology, Indian Institute of Technology-Madras (IIT-M), Chennai 600036, Tamil Nadu, India

*Corresponding Author

Anindita Bhadra (abhadra@iiserkol.ac.in)

**Abstract**

Animals employ strategic decision-making while carefully weighing nutritional benefits against the risks presented by aversive or harmful stimuli in their natural environment, to maximize foraging efficiency, In India, free-ranging dogs subsist predominantly on human-generated waste, where they often encounter food contaminated with unpalatable or noxious substances such as lemon juice while scavenging. The strategies these dogs use to navigate such challenges remain poorly understood, yet are critical for understanding their ecological adaptability and survival in human-dominated environments. A total of 156 randomly encountered free-ranging adult dogs were tested across 15 sites in Nadia district, West Bengal. Each individual was exposed to a single food source containing chicken placed in either lemon juice, diluted lemon solution, or water. All trials were video-recorded, and the behavioural sequences of the dogs, including sniffing, licking, eating, and food manipulation were coded and analysed to quantify strategic foraging responses under unpalatable conditions. They were found to use a flexible, multi-pronged strategy to manipulate the comparatively less palatable food option, and typically avoid the most unpalatable one, to maximize their acquiring options. Overall, this study revealed a hierarchically structured and context-dependent foraging strategy of free-ranging dogs, propelled by sensory evaluation, risk–reward balancing, and behavioural flexibility. These findings demonstrated how urban scavengers dynamically adapt to aversive conditions while scavenging, underscoring the cognitive mechanisms that support their survival in human-dominated environments.

**Introduction**

Foraging is not a simple search for food, but a series of tactical decisions. Confronted with a potential meal, an animal must first decide if the nutritional incentive is worth the handling cost, a trade-off between investment and avoidance. This initial evaluation forms the base of a cognitive hierarchy that dictates how animals interact with their resources. Decisions about what, when and where to explore for food, mates and shelter work as the driving force behind adaptive foraging in unpredictable environments (Dukas, 1998; Shettleworth, 2001). Animals operate as natural statisticians, making probabilistic judgments under ambiguity to balance effort and reward during foraging (Knill & Pouget, 2004; McNamara et al., 2006). These selections reflect Bayesian-like decision processes, where sensory cues and prior skills are integrated to update expectations dynamically (Budaev et al., 2019).

Free-ranging dogs (*Canis lupus familiaris*), with their ecological flexibility and reliance on human-dominated landscapes, constantly evaluate variable food sources from refuse to human provisioning, making them ideal for studying decision-making under uncertainty. This scavenging lifestyle exposes dogs to health risks, as contaminated food (e.g., with Streptococcus or Salmonella) can cause severe illness, highlighting adaptive trade-offs in foraging where efficiency must be balanced against aversive stimuli (Evans & Hooser, 2010; Coppock, 1983). Animals display defensive motivations (e.g., avoidance of toxins or predators) beside appetitive drives (Konorski, 1967; Lang et al., 1990), with decisions moulded by the reliability of risk cues like chemical or mechanical signals (Weissburg et al., 2014; Derby & Sorensen, 2008). For instance, prey species avoid predator exudates (Kats & Dill, 1998) or conspecific alarm cues (Chivers et al., 2002), while mechanosensory threats (e.g., predator contact) trigger rapid evasion (Frost et al., 1998). Free-ranging dogs, like other scavengers, must thus assess foraging gains against aversive outcomes, such as ingesting spoiled food or encountering harmful substances, highlighting the interplay between risk perception and adaptive decision-making (Rushen, 1996; Fraser & Matthews, 1997).

Research on dogs' cognition shows they can compare and select larger quantities, even when food is hidden, demonstrating mental representation in their interactions with physical and social environments (Csányi et al., 2001; Osthaus et al., 2003a,b; Ward & Smuts, 2006; Banerjee & Bhadra, 2019). The win-stay, lose-shift strategy, seen across species, promotes adaptability by repeating rewarded behaviours and shifting after unrewarded ones (Levine, 1959; Dember & Fowler, 1958; Zentall et al., 1990; Imhof et al., 2007; Badyna, 2024).

Decision-making in foraging and avoidance provides insights into survival strategies, as organisms often evade contaminated areas even at sub-lethal levels, reshaping ecosystems and serving as sensitive indicators of ecological risk (Moreira-Santos et al., 2008). Yet avoidance behaviours remain underrepresented in assessments, partly due to reliance on simplified two-option tests, highlighting the need for more ecologically relevant designs. Animal decision-making in foraging, social dynamics, and environmental navigation reflects complex interactions between cognitive strategies, ecological conditions, and evolutionary pressures (Budaev et al., 2019, Sih & Del Giudice, 2012). Free-ranging dogs (FRDs) in India rely heavily on human garbage, often consuming carbohydrate-rich, meat-scented foods (Bhadra et al., 2016a; R. Sarkar et al., 2019). Their foraging strategies are shaped by olfactory-driven food detection (Ruzicka & Conover, 2012) and risk assessment, akin to foxes that adjust behaviour to environmental threats (Berger-Tal et al., 2009; Vanak et al., 2009). FRDs prioritize protein-rich resources when available, following an evolved "Rule of Thumb" for efficient scavenging (Sarkar et al., 2019; Bhadra et al., 2016), while balancing predator avoidance with energy needs (Brown & Kotler, 2007; Haswell et al., 2017). However, their reliance on human waste exposes them to challenges like lemon-contaminated food. Unlike pet dogs, FRDs likely cannot afford to reject tainted food due to competition and scarcity (Pal et al., 2025a; Woodford & Griffith, 2012; Bhattacharjee et al., 2017). While an innate aversion to citrus might exist, survival pressures may override this constraint, forcing FRDs to consume lemon-contaminated food despite potential dislike. Recent studies reveal, however, that FRDs exhibit a clear aversion to high concentrations of lemon juice, adjusting their feeding strategies based on the specific lemon component, demonstrating their sensitivity to taste and adaptive foraging behaviour (Pal et al., 2025a). This selectivity is particularly prominent in adult dogs, who refine their strategies by preferring lower lemon concentrations and making more deliberate judgements when encountering unpalatable food (Pal, 2022: Pal et al., 2025a). Juveniles, in contrast, display less discriminatory foraging, suggesting that experience and cognitive maturity are critical in elevating feeding success (Pal et al., 2025b). This developmental trajectory highlights their flexibility: combining risk-sensitive foraging (Fortin et al., 2009) with opportunistic resource use (Deygout et al., 2010). Their success in urban ecosystems stems from adaptive decision-making, where survival pressures override innate aversions, and learning optimises scavenging efficiency. In this study we use lemon juice because it is a common, naturally acidic, and aversive contaminant frequently encountered by Indian free-ranging dogs in garbage, creating an ecologically relevant scenario to study their cost-benefit foraging decisions.

The objective of this study was to investigate the strategies and behavioural adaptations of Indian free-ranging dogs while scavenging in an unpalatable medium, specifically focusing on their interactions with acidic lemon juice. We aimed to identify how these dogs navigate challenges posed by unpalatable substances and the methods they employ to extract palatable food from such environments.

**Materials and Methods**

**Study Sites**

The experiment was conducted in 15 different locations within the Nadia district (22.9747° N, 88.4337° E) of West Bengal (Supplementary figure 1). The experiments were conducted in two time slots: 06:00 -12:00h and 15:00 - 20:00h.

**Selection and Identification of Dogs**

Only adult free-ranging dogs (based on size and genital structures) were used for this experiment. All tests were conducted on dogs that appeared to be healthy, i.e., did not have any obvious signs of disease or injury, and those that participated in the experiment by their choice. The gender of each dog was recorded during the experiment by observing their genitalia. Anthropogenic disturbance in the experimental areas was quantified as 'human flux' to evaluate fluctuations in dog behaviour in relation to human flux (Bhattacharjee et al., 2021). To estimate human flux, three one-minute videos were recorded at the experimental location, prior to and following each trial. Anthropogenic movement (people and vehicles) visible in the videos were counted, and the average of these counts was used as a measure of human flux. Random sampling was employed, primarily focusing on solitary adult dogs. In instances where dogs were observed in groups of two or three, one individual was randomly selected and lured away from the group to minimise external disturbances and reduce potential influences on the focal dog's behaviour.

**Materials**

The materials required for the experiment were gathered; these were fresh lemons, fresh chicken pieces (approximately 15 grams each), plastic gloves, and biodegradable paper bowls with a maximum capacity of 150 ml. A phone camera was used to record videos and take photos during the experiment. In order to ensure accuracy, fresh white biodegradable paper bowls, gloves, solutions, and chicken pieces were utilized for each trial. Lemon juice was extracted using a squeezer. We prepared three different setups for the experiment, each containing a single bowl. One bowl held 20 ml of 100% lemon juice with a 15 grams chicken piece, another contained 20 ml of 25% lemon juice solution in distilled water with a 15 grams chicken piece, and the third contained 20 ml of distilled water and a 15 grams chicken piece. For each trial, one of these setups was randomly selected and placed in front of the dog. This random selection ensured that each dog was exposed to only one type of food source during its trial. These distinct food sources were used to investigate how dogs respond to different types of food in unpalatable environments. Each trial, considered as a single phase of the experiment, involved only one food source per dog.

**Methodology**

A total of 156 dogs were randomly tested from 15 different locations for this experiment. Each dog participated in only one trial lasting approximately 90 seconds. The setup was positioned about 1-1.5 m in front of each dog's face to allow them to approach it freely. During the experiment, the experimenter stood approximately 1.5-2 m away from the setup with their hands open and arms straight, ensuring no eye contact with the focal dog. After 90 seconds, the setup was removed, and all materials used in that trial, including paper bowls, leftover solutions, and chicken pieces were discarded. New bowls, chicken pieces, and solutions were utilised for each subsequent trial (Fig. 1). A representative video of the experimental protocol is provided in Supplementary Video 1.

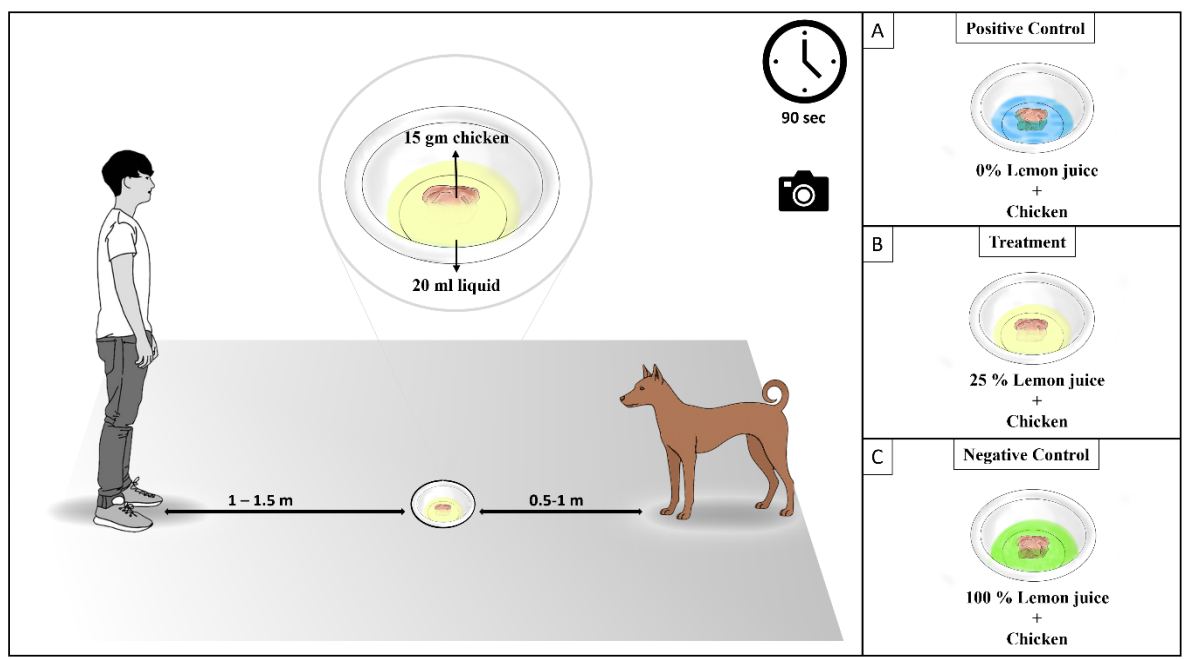

**Fig.1** Diagram representing the experimental setup used and the three distinct conditions presented to the dogs.

**Behavioural analysis**

Each instance of the focal dog's interaction with the chicken piece was documented, including the sequence of actions, the time taken for each behaviour, and the overall success in food consumption. The observed behaviours were classified into distinct categories based on their nature and function, including sniffing, licking, eating, food manipulation, and strategic modifications of food placement. The complete list of behaviours is provided in the ethogram (Table 1). The interactions were recorded from the moment the food bowl was placed until the completion of eating or the termination of engagement.

The first observed behaviour (e.g., sniffing or licking) was noted, followed by subsequent behaviours that contributed to the overall strategy. Sequential behavioural patterns were identified, providing insights into how dogs approached food-related challenges. The behavioural strategies employed by the dogs were varied; some engaged in direct consumption behaviours such as sniffing, licking, and eating, while others exhibited more complex manipulative strategies, including nudging the bowl, upturning it, or using their forelegs to access the food. Additionally, some dogs carried food away from the bowl before consuming it, while others engaged in behaviours like rubbing or shaking the food prior to eating. Instances of failed attempts, such as unsuccessful grabs for food, were also documented, highlighting the challenges dogs faced in manipulating food before successfully consuming it.

To assess the impact of external factors, behaviours were analysed in relation to experimental conditions, time of day, and location. Dogs' responses were compared across different conditions to identify significant variations in their strategic approaches. The study also examined the success rate of food consumption based on the behavioural strategies employed. Some dogs completed the eating process in a single sequence, while others required multiple attempts, incorporating various strategies before achieving successful consumption. The overall time taken for successful eating from the initial sniffing to the final bite was analysed to understand the efficiency of different behavioural approaches. Furthermore, post-consumption behaviours, such as sniffing or licking the empty bowl, were recorded to assess whether dogs exhibited any lingering interest in the food or its remnants. The presence of such behaviours indicated varying degrees of engagement with the food beyond direct consumption.

In order to effectively document these strategies, each behaviour was assigned a unique code based on predefined definitions. The order and pattern of strategic behaviours (SB1, SB2, SB3) were analysed to determine common response trends and variations across individuals and experimental conditions. The duration of each behaviour was measured in milliseconds, allowing for precise quantification of behavioural durations and transitions. Inter-rater reliability for behavioural coding was assessed, and the results are provided in Supplementary Table 3.

| Name of the strategy | Behavioural Code | Definition |
|---|---|---|
| Upturning bowl by using teeth | UB | The act of grasping a bowl with teeth and partially or fully flipping it over. The act of carrying the bowl of food and some spillage of liquid or food or both. |
| Head shake | SH | The rapid sideways movement of the head following the act of grabbing food in the mouth. |
| Head shake without food | SHNF | The rapid sideways movement of the head without food in the mouth. |
| Placing food on the ground | PG | The act of placing food on the ground in a courteous manner, ensuring it's not dropped but set down gently. |
| Drop food on the ground | DG | Dropping food from a height onto the ground, not employing a gentle movement. |
| Rubbing food on the ground | RB | Rub the food against the ground using teeth, mouth, or snout. Dogs can drag the chicken piece using teeth/mouth on the ground. |
| Using foreleg | FL (FLL, FLR, FLB) | The action of manipulating food using one or more forelegs (left, right, or both) to free food or tamper the food or tamper the bowl. |
| Multiple lick | ML | The action of repetitively licking using the tongue, occurring more than once, typically two or more times. |
| Nudging bowl | NB | A behaviour where dog nudges the bowl containing food, using its snout, typically resulting in a mild movement or displacement of the bowl. |
| Upturn bowl by nudging | UBNB | The act of overturning a bowl by applying force through a gentle nudge, resulting in a complete or partial reversal of its position rather than just a mild displacement. |
| Only Chewing | CW | The action of a dog using its mouth to bite, gnaw, or chew on various objects such as toys, bones, or food items. Chewing but after that not eating the food. If |

| | | chewing and then eat the food then it will not be considered as chewing. |
|---|---|---|
| Eat | ET | The act of consuming food entirely, including consuming all parts together. If chewing and then after chewing eating the whole food, it will be counted as eating. |
| Partial eating | PE | Specific eating behaviour when instead of eating entire chicken piece, parts of the chicken piece is being eaten one at a time. If a chicken piece is being chewing and eating, it will be considered as PE. |
| Sniff | SN | The dog approaches the bowl and investigates it by sniffing from a distance of within 10cm. |
| Lick | LI | The act of a dog using its tongue to clean or consume food from a bowl. This behaviour excludes the use of the muzzle or snout. |
| Carrying food or bowl of food or bowl | CF | The act of picking of the chicken piece from the bowl and walking/ trotting/running away from the bowl or from the ground with chicken piece inside its mouth. It can be positional displacement of the food from the bowl. The act of carrying the bowl of food, bowl and place it on the ground. The act of carrying food. |
| Failed grab | FG | This category includes all the situations in which the dog tries to take up the chicken with its mouth but cannot manage to do so. This may entail making tries in the bowl or on the ground. It eventually drops the food several times on the ground or bowl. |
| Not visible | NV | When dog's activity is not detectable and visible from the decoder's point of view. |

**Table 1:** An ethogram of observed behaviours in free-ranging dogs during this experiment.

**Statistical analysis**

Behavioural and time-based data were analysed to assess strategic foraging patterns of free-ranging dogs across conditions A, B, and C. Frequencies of strategic behaviours (excluding sniffing) were clustered into high, medium, and low groups using K-means clustering to group these behaviours based on their occurrence rates. Within-cluster comparisons were conducted using Fisher's Exact Test, while conditions-wise differences were assessed with Chi-square or Fisher's Exact Test (for low counts), followed by Bonferroni-corrected post-hoc comparisons. Latency measures, from bowl placement to first sniff and first lick, as well as total consumption time, were calculated, tested for normality using Shapiro-Wilk tests, and compared across conditions with Kruskal-Wallis tests and Dunn's post-hoc corrections.

Time-to-event data (latency to sniff or eat) were analysed with Kaplan-Meier survival curves, with phase comparisons via log-rank tests. Cox proportional hazards models were fitted to examine the effects of Condition and Sex, with proportional hazards assumptions verified using Schoenfeld residuals. Adjusted survival curves and hazard ratios were visualized using ggadjustedcurves and ggforest.

Directed behavioural sequence networks were constructed for each condition (igraph, ggraph), with nodes representing behaviours and edges representing transitions. Sequences were cleaned and converted into edge lists. Node-level metrics (degree, betweenness, closeness, PageRank) quantified behavioural influence, and network-level metrics (density, reciprocity, centralization, connectivity, strongly connected components, characteristic path length) assessed overall structure. Shortest path matrices evaluated behavioural reachability, and changes in degree between consecutive conditions (A→B, B→C) highlighted shifts in behavioural influence. Pairwise transition counts were computed and network non-randomness was tested by comparing observed structures with 1,000 random networks using clustering coefficient Z-scores.

All analyses were performed in R (v4.2.2) and MATLAB (R2023a) using betareg, ggeffects, igraph, ggraph, ggplot2, dplyr, tidyr, patchwork, survminer, and ggforce, providing a comprehensive assessment of conditions- and sex-dependent variations in strategic foraging behaviour.

# Results

## 1. Patterns of strategic foraging behaviours across experimental conditions?

Out of sixteen observed behaviours, eight showed a statistically significant difference in frequency across the experimental conditions (Adjusted $P < 0.05$), indicating that the free-ranging dogs employed condition-dependent strategic adaptations to the unpalatable medium. Significant differences were consistently found in behaviours associated with direct interaction and manipulation of the food: Eating (ET) and Placing on the Ground (PG), along with behaviours related to aversion and assessment, such as Multiple Licking (ML), Shaking Head (SH), and Rubbing on the Ground (RB). Notably, foraging success or Eating (ET) was the most variable behaviour, differing significantly across all three pairwise comparisons (A vs. B, A vs. C, and B vs. C), underscoring the strong impact of the varying unpalatability level on the final outcome. Manipulative strategies like Placing on the Ground (PG) showed significant variation between Condition A vs. B and B vs. C, demonstrating a shift in effort investment as the challenge changed. Aversive behaviours like Multiple Licking (ML) were significantly different in comparison with Condition C (A vs. C, B vs. C), suggesting that the highest level of unpalatability triggered a more pronounced rejection response (Supplementary Table 1). These results quantify the dogs' decision to invest effort or avoid and confirm that their scavenging tactics are dynamically adjusted in response to the perceived cost of overcoming the acidic contamination (Fig. 2).

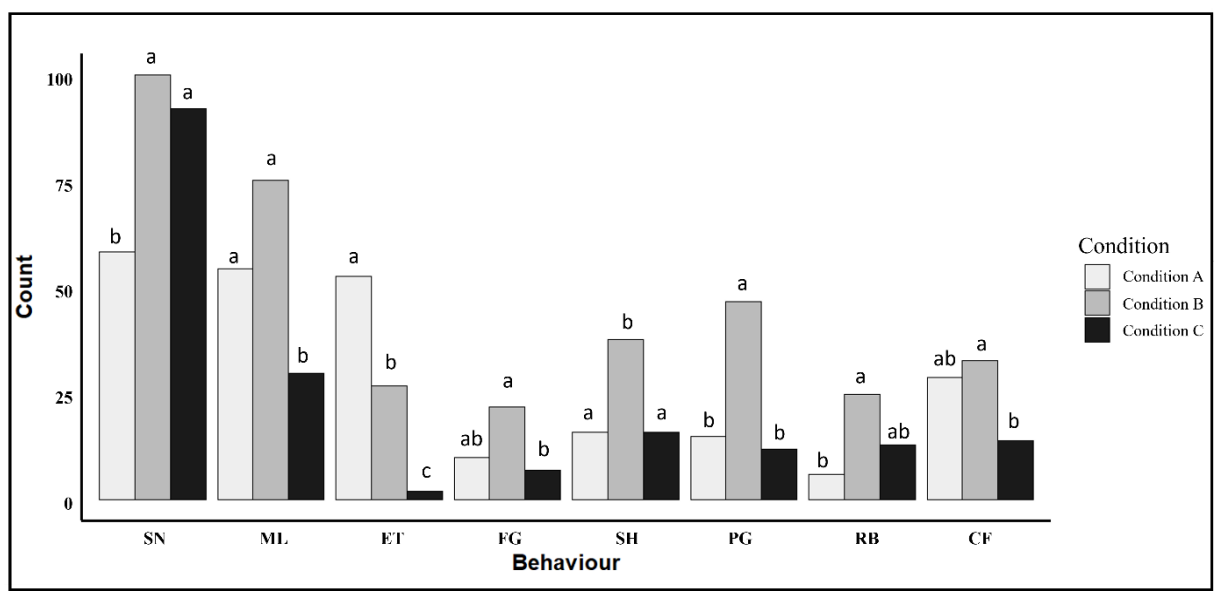

**Figure 2:** A bar graph showing the frequency of observed behaviours across experimental conditions. The letters denote differences across conditions for each behaviour. All comparisons are within a behaviour type, across conditions.

## 2. Time investment in strategic foraging behaviours by free-ranging dogs across the experimental conditions

The beta regression analysis comparing Condition B and Condition C to the reference Condition A for various strategic behaviours revealed several significant findings.

For Sniff (SN), Condition C demonstrated a significant positive effect (Estimate = 1.265, $p < 0.001$), while Condition B did not show a significant effect. No significant effects were observed for Multiple Lick (ML) in either condition. Eat (ET) exhibited significant negative effects in both Condition B (Estimate = -0.938, $p < 0.001$) and Condition C (Estimate = -2.026, $p = 0.003$). Similarly, Failed Grab (FG) showed significant negative effects in both conditions, with Condition B (Estimate = -0.970, $p < 0.001$) and Condition C (Estimate = -1.024, $p = 0.005$).

For Head Shake (SH), a significant negative effect was found in Condition C (Estimate = -0.889, $p = 0.003$), while Condition B was not significant. Placing Food on the Ground (PG) demonstrated significant negative effects in both conditions, with Condition B (Estimate = -0.671, $p = 0.004$) and Condition C (Estimate = -1.270, $p < 0.001$). In the case of Lick (LI), the intercept was significant (Estimate = -1.384, $p = 0.021$), but neither condition showed significant effects. Rubbing Food on the Ground (RB) was significantly affected in Condition C (Estimate = -0.617, $p = 0.045$), whereas Condition B was not significant (Fig. 3).

Carrying Food or Bowl (CF) showed significant negative effects in both Condition B (Estimate = -1.053, $p < 0.001$) and Condition C (Estimate = -0.947, $p < 0.001$). For Upturn Bowl by Nudging (UBNB), the intercept was significant (Estimate = -1.711, $p = 0.035$), but neither condition had a significant effect. Lastly, for Nudging Bowl (NB), both the intercept (Estimate = -3.317, $p = 0.001$) and the phi coefficient (Estimate = 10.885, $p = 0.027$) were significant, although neither condition showed a significant effect (Supplementary Table 2).

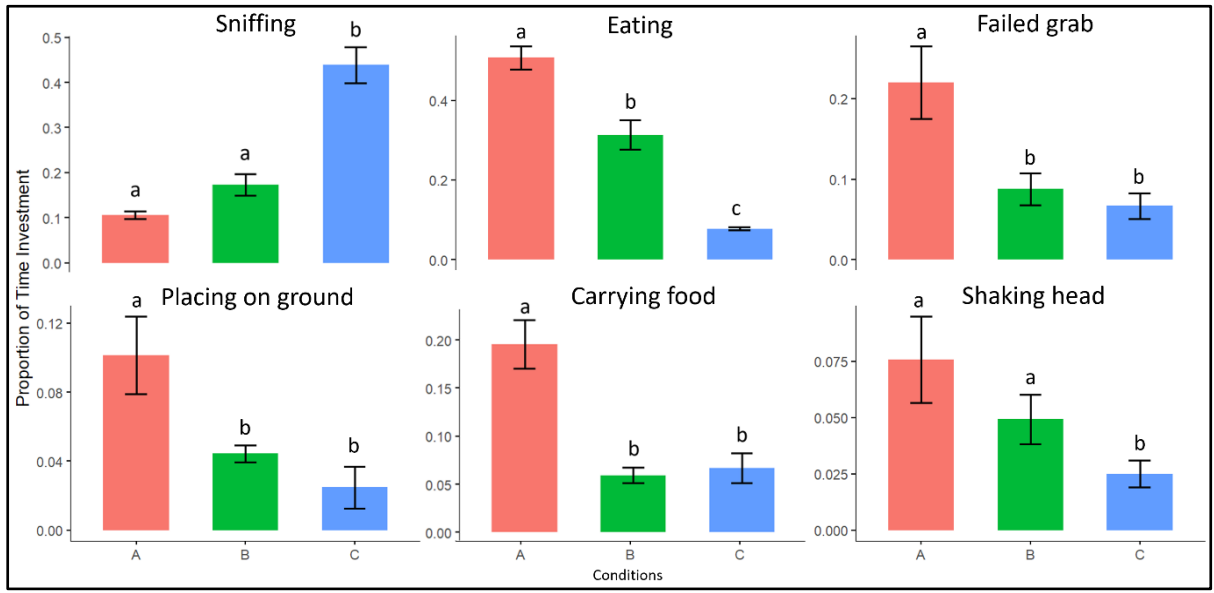

**Figure 3**: Bar graphs showing the mean ± s.d. of the proportion of time invested by free-ranging dogs in six behaviours across the three experimental conditions.

In summary, significant effects were predominantly observed for Condition C across multiple strategic behaviours, indicating its influence on time investment, while Condition B exhibited limited significance. Notably, gender did not have a significant impact in any of the models.

## 3. The effect of lemon juice concentration

### (i) Total time taken and likelihood of food consumption

In order to assess if the presence of lemon juice influenced the time taken by free-ranging dogs to complete food consumption, we measured the total time from the first sniff to the completion of eating across three experimental conditions: Condition A (0% lemon juice), Condition B (25%), and Condition C (100%). A Kruskal-Wallis rank sum test revealed a significant difference in total time taken across the conditions ($\chi^2$ = 25.274, df = 2, p = 3.25e-06). Post-hoc pairwise comparisons using Dunn's test with Bonferroni correction showed that dogs in Condition A took significantly less time compared to Condition B (p < 0.0001). However, no significant difference was observed between Condition A and Condition C (p = 0.09), or between Condition B and Condition C (p = 1.00). These results suggest that lemon juice exposure at 25% significantly delayed food consumption, but further increase to 100% did not lead to a statistically distinguishable increase in time taken, potentially due to near-complete avoidance in Condition C (Fig. 4).

Further, we examined the likelihood and timing of food consumption using a Kaplan-Meier survival analysis, where the event of interest was the completion of eating. Median survival time (i.e., time from first sniff to complete eating) was shortest in Condition A at 13.25 seconds (95% CI: 11.25–16.79), and increased substantially in Condition B to 85.62 seconds (95% CI: 52.00–NA). In Condition C, the median survival time could not be estimated because only 2 of 50 dogs completed eating within the 90-second observation window, indicating extreme aversion.

A log-rank test confirmed significant differences in the survival distributions across conditions ($\chi^2 = 160$, df = 2, $p < 2e-16$). The Cox proportional hazards model, with Condition A (0% lemon juice) as the reference category, revealed that dogs in Condition B were 87% less likely to complete food consumption (HR = 0.129, 95% CI: 0.076–0.220, $p < 0.001$), and in Condition C, the likelihood further dropped by 99.3% (HR = 0.0074, 95% CI: 0.0017–0.031, $p < 0.001$), relative to Condition A. These findings indicate a dose-dependent aversive effect of lemon juice on feeding, with dogs being significantly less likely to complete eating as citrus concentration increased (Supplementary information 1).

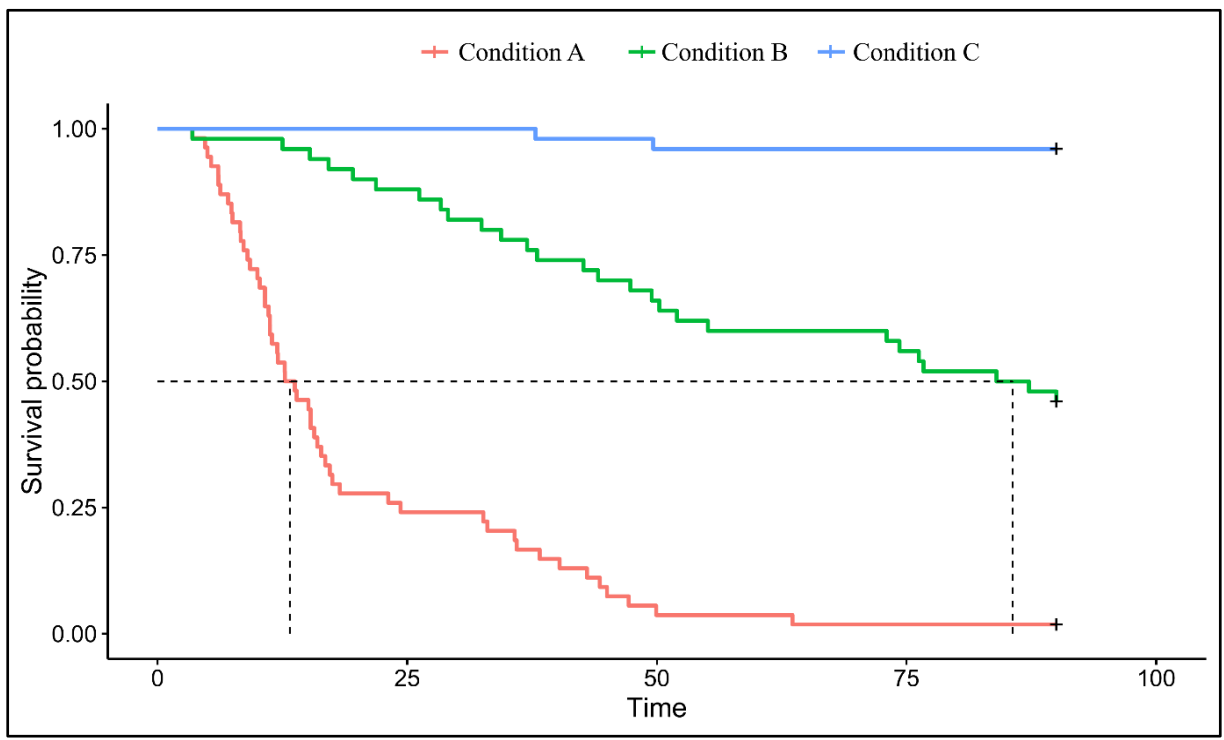

**Figure 4:** A survivorship plot showing the probability of food remaining over time across different experimental conditions.

Although gender data were partially recorded, inclusion of gender as a covariate in the Cox model revealed no significant effect (HR = 1.421, 95% CI: 0.906–2.227, p = 0.126), suggesting that male and female dogs responded similarly to citrus exposure. The Cox model showed good predictive performance, with an AIC of 613.45 and a concordance index (C-index) of 0.85 (Supplementary figure 2).

Together, these results demonstrate that increasing lemon juice concentration prolongs the time taken to eat and significantly reduces the likelihood of food consumption, particularly under 100% lemon juice exposure, where dogs largely avoid eating altogether.

**(ii) Latency and the time taken for the first lick after the first sniff**

We initially used the Kruskal-Wallis rank sum test to check whether increasing lemon juice concentrations influenced the latency to approach the food (defined as the time from release to the first sniff). The analysis revealed no significant difference in latency across experimental conditions ($\chi^2$ = 0.703, df = 2, p = 0.704), indicating that the initial motivation or approach tendency of the dogs remained consistent regardless of the lemon juice concentration. However, given that latency inherently reflects time-to-event data and some dogs may never approach within the observation period, a survival analysis framework would be more appropriate to assess this metric in future analyses. The current test may underestimate nuanced treatment effects where censored data is involved (Supplementary information 2).

In contrast, the interval between first sniff and first lick, which reflects hesitation or sensory evaluation before initial contact with food, did show a significant effect of treatment. A Kruskal-Wallis test revealed a highly significant difference across conditions ($\chi^2$ = 35.264, df = 2, p = 2.2e-08). Post-hoc Dunn's tests with Bonferroni correction confirmed that dogs in Condition B (25% lemon juice) and Condition C (100%) showed significantly longer intervals compared to Condition A (0%) (p < 0.0001 for both comparisons). However, the difference between Condition B and Condition C was not significant (p = 0.364), suggesting that the presence of lemon juice, even at moderate concentration, introduced a clear delay between olfactory investigation and actual food contact, but increasing the concentration from 25% to 100% did not further prolong this interval (Supplementary information 3).

## 4. The effect of experimental conditions on the behavioural networks

Network analysis revealed distinct structural differences in behavioural interactions across the three experimental conditions. Condition B (Treatment) exhibited the highest network density (1.50) and clustering coefficient (0.63), along with the shortest average path length (1.82) and diameter (3), suggesting more interconnected and efficient behavioural exchanges compared to other conditions. In contrast, Condition C (Negative Control) exhibited the sparsest network (density = 0.77) and the lowest clustering (0.50), with longer path lengths (2.01), indicating fragmented behavioural connectivity. Condition A (Positive Control) displayed intermediate properties (density = 1.13, clustering = 0.56), but its diameter (4) matched Condition C, reflecting comparable network breadth.

Centrality metrics highlighted condition-dependent shifts in key behaviours: ML (multiple lick) dominated Condition A (degree = 109, betweenness = 73.7), while Condition B saw increased prominence of SN (sniffing; degree = 139, betweenness = 78.9) and PG (pawing/grasping; degree = 94). Condition C exhibited a reversal to SN-centric interactions (degree = 98, betweenness = 56.2), though with reduced ML engagement (degree = 55). Notably, Condition B demonstrated elevated betweenness for cooperative behaviours (e.g., RB [rolling/balling] = 18.1) compared to Condition A (RB = 3.4), suggesting enhanced role differentiation during treatment. These findings collectively indicate that Condition B's network topology fostered tighter behavioural integration, whereas Conditions A and C displayed either balanced or decentralise d interaction patterns, respectively.

All three experimental conditions showed significantly non-random behavioural interaction structures compared to their corresponding random networks (all $p < 0.0001$). The observed clustering coefficients were markedly higher than expected by chance, as indicated by the high Z-scores in each condition: Condition A ($z = 7.28$, clustering = 0.9119), Condition B ($z = 16.86$, clustering = 1.3097), and Condition C ($z = 13.67$, clustering = 1.2808). Among these, the Treatment Condition (B) exhibited the strongest deviation from randomness ($z = 16.86$), indicating the highest level of behavioral organization and coordinated interaction among the dogs. The consistency of these effects across conditions implies that, while network density and centrality varied (as previously reported), the tendency for behaviours to cluster was a universal feature of the experimental paradigm, exceeding chance expectations. This structural property may reflect underlying cognitive or social constraints shaping behavioural sequences (Supplementary information 4) (Supplementary figure 3, 4).

## 5. Behavioural sequences during successful foraging events

A total of 27 behavioural sequences ending with eating ("ET") were analysed, comprising 250 transitions across 14 distinct behavioural states (Supplementary Table 4). The empirical Markov transition matrix revealed that dogs' behavioural organization in successful trials was highly structured, with a few transitions dominating the network. The most frequent behavioural shifts were from SN → ML (n = 22), SH → PG (n = 14), and CF → PG (n = 13), suggesting that movement toward the multiple lick (ML) and placing on the ground (PG) were central components of successful foraging strategies. Notably, transitions involving SH → ET (n = 10) and RB → ET (n = 5) indicated that rubbing on the ground (RB) actions often directly preceded food acquisition, reflecting sensory validation and rapid decision closure. Overall, ML, PG, SH, and CF emerged as key intermediary nodes linking exploration to goal achievement, highlighting a non-random, condition-dependent behavioural sequence architecture associated with success in the garbage-phase task. To understand the optimal sequential path from the first exploratory behaviour "SN" to the acquisition behaviour "ET", we checked the dominant transitions (probability >/= 0.2) (Fig. 7) and plotted in a network plot minimising noise from the rare transitions. The complete behavioural transition network is provided in the Supplementary Material (Supplementary figure 5).

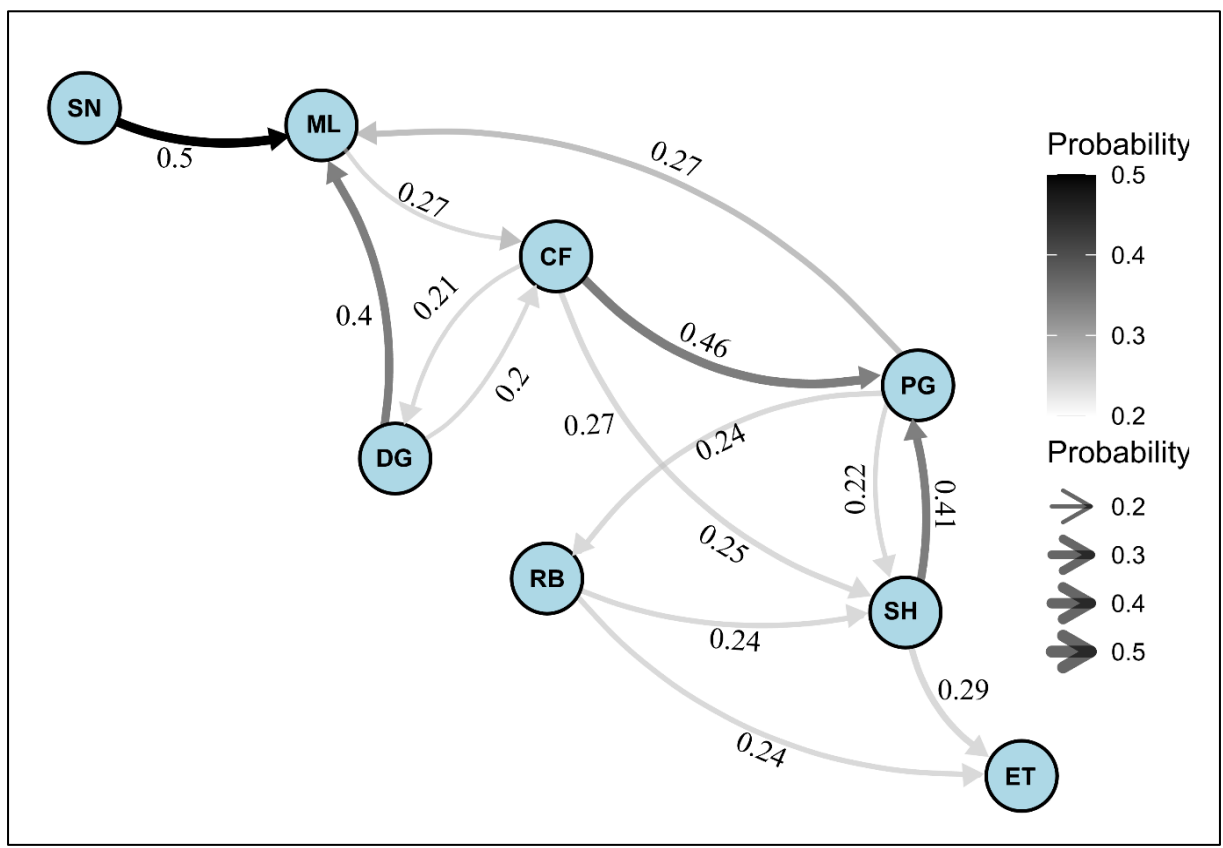

**Figure 5:** Behavioural transition network from Sniffing (SN) towards Eating (ET). Nodes represent behaviours, and directed edges indicate transitions with probabilities ≥ 0.2. Edge thickness and colour intensity are proportional to transition probability. Only high-probability connections are displayed to emphasize stable behavioural routes and minimize noise from rare transitions.

**Discussion**

Controlled field experiments were conducted on solitary adult free-ranging dogs, by presenting them a palatable food source under conditions with increasing aversive strength. The primary aim was to understand if and how the dogs use various strategies to obtain the food from different levels of unpalatable backgrounds.

Our experiments revealed a flexible, context-dependent foraging strategy, with ample variation in the use of different behaviours to obtain the food across different conditions. We recorded sixteen different behaviours that we label as strategizing behaviours, and eight of these varied considerably in the frequency of occurrence across conditions. This behavioural flexibility leads to adaptive foraging, in which animals assess the benefits of eating against the costs of interacting with unpleasant stimuli. Their foraging strategy choices seems to be guided by dynamic cost-benefit analyses, with tactics influenced by the perceived difficulty of accessing palatable food, consistent with Optimal Foraging Theory (Pyke, 1984; Stephens & Krebs, 1986). This flexibility is further highlighted by time investment patterns. During the treatment (25% lemon juice) and negative control (100% lemon juice) conditions, there was a noticeable decrease in eating, with a corresponding increase in sniffing, which suggests that the reduced gain was quickly recognized. Strategic, assessment-driven foraging, in which dogs actively regulate their temporal budget by reallocating effort toward problem-solving, is exemplified by this transition from direct eating to gathering information under uncertainty.

This study shows that lemon juice wields a clear, dose-dependent aversive effect, indicating the existence of a strong threshold beyond which feeding motivation sharply declines. Once the sourness crosses this limit, dogs tend to minimize their energy investment in procuring the food item, implying a cost benefit decision making rather than a simple preference. This act aligns with risk–reward trade-off models of decision-making (Kacelnik & Bateson, 1996). Similar context-dependent foraging shifts have been observed in hummingbirds and jays (Hurly & Oseen, 1999; Shafir et al., 2002). For urban scavengers frequently exposed to spoiled or contaminated food, such rapid risk–reward evaluation is crucial for survival. These findings suggest that citrus-based repellents offer a humane way to reduce human–dog conflict while revealing how urban scavengers integrate sensory aversion into decision-making for coexistence in rapidly urbanising environments.

Dogs approached food with ease but paused after close olfactory investigation, according to behavioural patterns that showed a two-stage foraging process, consistent with cue-based sensory foraging frameworks (Stoddart, 1980). Lemon juice acted as a sensory cue and caused substantial resistance subsequently, but it didn't hinder approach. Olfactory tolerances are crucial for risk-averse foraging since foraging decisions are contingent upon close-range olfactory assessments. The differential response upon sniffing to the 25% and 100% lemon juice contamination underscores the role of olfactory cues in foraging decision-making in the dogs.

Network analysis depicted clear condition-dependent reorganization of behavioural connectivity, indicating that dogs adopt their foraging behavioural structure based on the level of lemon concentration. Although food avoidance was high under 100% lemon concentration, the residual behavioural sequences that occurred were non-randomly organized, suggesting that even in highly aversive contexts, dogs' behavioural responses remain structured rather than chaotic. Similar context-dependent flexibility has been documented in corvids and primates, reflecting convergent cognitive evolution (Emery & Clayton, 2004; Laumer et al. 2019). Their foraging strategies appeared to follow organized decision-making processes rather than random activities, reminiscent of hierarchical models of sequential decision-making (Daw et al., 2005; Balleine & Dickinson, 1998). Exploratory and manipulative behaviours emerged as the key to foraging success, with the transition network demonstrating non-random, directed behavioural pathways having high-probability links between exploratory, manipulative, and consummatory acts. This structured foraging pattern and organized progression from assessment to acquisition, and the differential responses to the three tested conditions, highlight the flexibility in the foraging strategy of the free-ranging dogs.

 The study revealed a complex picture of behavioural plasticity, strategic decision-making, and adaptive problem-solving abilities in the context of foraging in Indian free-ranging dogs.  The structure, timing and the sequence of behaviours in response to the differential aversive stimulus elucidate a sophisticated cognitive architecture beyond a simple stimulus-response model. Free-ranging dogs recruit a hierarchically structured, dynamically reconfigurable foraging strategy, demonstrated through the differential reallocation of time, the dissociation between approach and consumption, and the dose-dependent aversive response to lemon juice. The emergence of distinct behavioural patterns depicts the dogs' cognitive flexibility, enabling them to switch between efficient exploitation and focused investigation based on environmental conditions. This behavioural plasticity that reinforces their success as urban scavengers (Lowry et al., 2013) and reflects the animal tendency to assess and balance risk and reward when exploiting unpredictable or contaminated resources. Building on this, future studies could examine the consistency of strategic foraging behaviours in individual free-ranging dogs and investigate how varying concentrations of sour substances further modulate their decision-making strategies. By integrating macro-level behavioural organization with micro-level responses to aversive stimuli, this study establishes free-ranging dogs as a valuable model for understanding the cognitive mechanisms of behavioural flexibility, risk management, and problem-solving in urban-adapted species worldwide.

**Ethical Statement**

The study design did not violate the Animal Ethics regulations of the Government of India (Prevention of Cruelty to Animals Act 1960, Amendment 1982). The protocol for the experiment was approved by the IISER Kolkata Animal Ethics Committee.

**Conflict of interest statement**

All the authors have read and agree with this version of the manuscript. The authors declare no conflict of interest.

**Authors' contributions**

TSP and AB conceptualised and designed the study. TSP, HG, and SN conducted fieldwork. TSP, SN and HG were involved in data decoding, and TSP, AM, and SB performed data analysis. SB also contributed to the preparation of illustrations. RS provided support and guidance during the manuscript writing and reviewing process. AB supervised the project, offered continuous input, and critically reviewed the manuscript. SN and HG contributed equally as joint second authors.

## Acknowledgements

We thank all members of the BEL Lab for their valuable insights and support throughout the study. We are especially grateful to Imran Mondal for his assistance during fieldwork. We also extend our sincere thanks to Aesha Lahiri and Epil Mandi for their valuable contribution to this project.

## Funding

TSP was supported by a PhD fellowship of the University Grants Commission (UGC) India. This study was supported by IISER Kolkata ARF and the Janaki Ammal National Women Bioscientist Award (BT/ HRD/NBA-NWB/39/2020-21 (YC-1)) of the Department of Biotechnology, India. SN was supported by the INSPIRE PhD fellowship, Department of Science and Technology, India and RS was supported by IISER Kolkata's PhD fellowship.

## Supplementary data

Will be made available on publication of the manuscript after peer review.

## References

1. Badyna, J. (2024). *The role of task events, environmental uncertainty, and striatal activity in modulating decision strategies* (Doctoral dissertation, Carnegie Mellon University). Carnegie Mellon University.
2. Balleine, B. W., & Dickinson, A. (1998). Goal-directed instrumental action: contingency and incentive learning and their cortical substrates. Neuropharmacology, 37(4-5), 407-419.
3. Banerjee, A., & Bhadra, A. (2019). The more the merrier. Current Science, 117(6), 1095-1100.

4. Berger-Tal O, Mukherjee S, Kotler BP, Brown JS (2009) Look before you leap: is risk of injury a foraging cost? Behav Ecol Sociobiol 63:1821–1827. https://doi.org/10.1007/s00265-009-0809-3
5. Bhadra, A., Bhattacharjee, D., Paul, M., Singh, A., Gade, P. R., Shrestha, P., & Bhadra, A. (2016a). The meat of the matter: a rule of thumb for scavenging dogs? Ethology Ecology & Evolution, 28(4), 427–440. https://doi.org/10.1080/03949370.2015.1076526
6. Bhattacharjee, D., N, N. D., Gupta, S., Sau, S., Sarkar, R., Biswas, A., ... & Bhadra, A. (2017). Free-ranging dogs show age related plasticity in their ability to follow human pointing. PloS one, 12(7), e0180643.
7. Bhattacharjee, D., Sau, S., Das, J., & Bhadra, A. (2017). Free-ranging dogs prefer petting over food in repeated interactions with unfamiliar humans. Journal of Experimental Biology, 220(24), 4654–4660. https://doi.org/10.1242/jeb.166371
8. Brown, J. S., & Kotler, B. P. (2007). Foraging and the ecology of fear. Foraging: Behaviour and ecology, 437-482.
9. Budaev, S., Jørgensen, C., Mangel, M., Eliassen, S., & Giske, J. (2019). Decision-making from the animal perspective: Bridging ecology and subjective cognition. Frontiers in Ecology and Evolution, 7, 164. https://doi.org/10.3389/fevo.2019.00164
10. Chivers, D., Mirza, R., & Johnston, J. (2002). Learned recognition of heterospecific alarm cues enhances survival during encounters with predators. Behaviour, 139(7), 929-938.
11. Coppock, C. E. (1983). Nutritional perspective on dairy cattle feeding systems.
12. Csa´nyi V, Topa´l J, Gacsi M, Sarkozi Z (2001) Distinguishing logic from association in the solution of an invisible displacement task by children (Homo sapiens) and dogs (Canis familiaris): using negation of disjunction. J Comp Psychol 115(3):219–226
13. Daw, N. D., Niv, Y., & Dayan, P. (2005). Uncertainty-based competition between prefrontal and dorsolateral striatal systems for behavioral control. Nature neuroscience, 8(12), 1704-1711.
14. Dember, W.N., & Fowler, H. (1958). Spontaneous alternation behaviour. *Psychological Bulletin, 55*, 412–428.
15. Derby, C. D., & Sorensen, P. W. (2008). Neural processing, perception, and behavioural responses to natural chemical stimuli by fish and crustaceans. Journal of chemical ecology, 34, 898-914.

16. Deygout, C., Gault, A., Duriez, O., Sarrazin, F., & Bessa-Gomes, C. (2010). Impact of food predictability on social facilitation by foraging scavengers. Behavioural Ecology, 21(6), 1131-1139.
17. Dukas, R. (Ed.). (1998). *Cognitive ecology: the evolutionary ecology of information processing and decision making*. University of Chicago Press.
18. Emery, N. J., & Clayton, N. S. (2004). The mentality of crows: convergent evolution of intelligence in corvids and apes. science, 306(5703), 1903-1907.
19. Evans, T. J., Hooser, S. B., & Second, E. (2010). 10.16-Comparative Gastrointestinal Toxicity. *McQueen CABT-CT. Oxford: Elsevier*, 195-206.
20. Fortin, D., Fortin, M. E., Beyer, H. L., Duchesne, T., Courant, S., & Dancose, K. (2009). Group-size-mediated habitat selection and group fusion–fission dynamics of bison under predation risk. Ecology, 90(9), 2480-2490.
21. Fraser, D., Matthews, L.R., 1997. Preference and motivation testing. In: Appleby, M.C., Hughes, B.O. (Eds.), Animal Welfare. CAB International, Wallingford, UK, pp. 159–173.
22. Frost, W. N., Brandon, C. L., & Mongeluzi, D. L. (1998). Sensitization of theTritoniaEscape Swim. Neurobiology of learning and memory, 69(2), 126-135.
23. Haswell, P. M., Kusak, J., & Hayward, M. W. (2017). Large carnivore impacts are context-dependent. Food Webs, 12, 3-13.
24. Hurly, T. A., & Oseen, M. D. (1999). Context-dependent, risk-sensitive foraging preferences in wild rufous hummingbirds. *Animal Behaviour*, *58*(1), 59-66.
25. Imhof, L. A., Fudenberg, D., & Nowak, M. A. (2007). Tit-for-tat or win-stay, lose-shift?. *Journal of theoretical biology*, *247*(3), 574-580.
26. Kacelnik, A., & Bateson, M. (1996). Risky theories—the effects of variance on foraging decisions. American zoologist, 36(4), 402-434.
27. Kats, L. B., & Dill, L. M. (1998). The scent of death: chemosensory assessment of predation risk by prey animals. Ecoscience, 5(3), 361-394.
28. Knill, D. C., & Pouget, A. (2004). The Bayesian brain: the role of uncertainty in neural coding and computation. *TRENDS in Neurosciences*, *27*(12), 712-719.
29. Konorski, J., 1967. Integrative Activity of the Brain: An Interdisciplinary Approach. University of Chicago Press, Chicago, IL
30. Lang, P.J., Bradley, M.M., Cuthbert, B.N., 1990. Emotion, attention, and the startle reflex. Psychol. Rev. 97, 377–395.

31. Laumer, I. B., Auersperg, A. M., Bugnyar, T., & Call, J. (2019). Orangutans (Pongo abelii) make flexible decisions relative to reward quality and tool functionality in a multi-dimensional tool-use task. PloS one, 14(2), e0211031.
32. Levine, M. (1959). A model of hypothesis behaviour in discrimination learning set. *Psychological review*, *66*(6), 353.
33. Levine, M. (1975). *A cognitive theory of learning: Research on hypothesis testing*. Routledge.
34. Lowry, H., Lill, A., & Wong, B. B. (2013). Behavioural responses of wildlife to urban environments. *Biological reviews*, *88*(3), 537-549.
35. McNamara, J. M., Green, R. F., & Olsson, O. (2006). Bayes' theorem and its applications in animal behaviour. *Oikos*, *112*(2), 243-251.
36. Moreira-Santos M, Donato C, Lopes I, Ribeiro R (2008) Avoidance tests with small fish: determination of the median avoidance concentration and of the lowest-observed-effect gradient. Environ Toxicol Chem 27:1575–1582
37. Osthaus B, Lea SEG, Slater AM (2003b) Training influences problem-solving abilities in dogs (Canis lupus familiaris). Proceedings of the Annual Meeting of British Society of Animal Science, York 103
38. Osthaus B, Slater AM, Lea SEG (2003a) Can dogs defy gravity? A comparison with the human infant and a non-human primate. Dev Sci 6(5):489–497
39. Pagé, D. D., & Dumas, C. (2003). Strategy planning in cats (Felis catus) in a progressive elimination task. *Journal of Comparative Psychology*, *117*(1), 53.
40. Pal, T. S. (2022). Understanding the preference of citrus food among free-ranging dogs (Master's thesis, Indian Institute of Science Education and Research Kolkata). IISER Kolkata Repository. http://eprints.iiserkol.ac.in/id/eprint/1337
41. Pal, T. S., Debnath, P., Biswas, S., & Bhadra, A. (2025). Lemons and Learning: Adult and Juvenile Free-Ranging Dogs Navigate Aversive Foraging Challenges Differently. https://doi.org/10.48550/arXiv.2504.08077
42. Pal, T. S., Nandi, S., Sarkar, R., & Bhadra, A. (2025). When Life Gives You Lemons, Squeeze Your Way Through: Understanding Citrus Avoidance Behaviour by Free-Ranging Dogs in India. Applied Animal Behaviour Science, 106682. https://doi.org/10.1016/j.applanim.2025.106682
43. Pyke, G. H. (1984). Optimal foraging theory: a critical review. Annual review of ecology and systematics, 15, 523-575.

44. Rushen, J., 1996. Using aversion learning techniques to assess the mental state, suffering, and welfare of farm animals. J. Anim. Sci. 74, 1990–1995.

45. Ruzicka RE, Conover MR (2012) Does weather or site characteristics infuence the ability of scavengers to locate food? Ethology 118:187–196. https://doi.org/10.1111/j.1439-0310.2011.01997.x

46. Sarkar, R., Sau, S., & Bhadra, A. (2019). Scavengers can be choosers: A study on food preference in free-ranging dogs. Applied Animal Behaviour Science, 216, 38–44. https://doi.org/10.1016/j.applanim.2019.04.012

47. Shafir, S., Waite, T. A., & Smith, B. H. (2002). Context-dependent violations of rational choice in honeybees (Apis mellifera) and gray jays (Perisoreus canadensis). *Behavioral Ecology and Sociobiology*, *51*(2), 180-187.

48. Shettleworth, S. J. (2001). Animal cognition and animal behaviour. *Animal behaviour*, *61*(2), 277-286.

49. Sih, A., & Del Giudice, M. (2012). Linking behavioural syndromes and cognition: A behavioural ecology perspective. Philosophical Transactions of the Royal Society B: Biological Sciences, 367(1603), 2762–2772. https://doi.org/10.1098/rstb.2012.0216

50. Stephens, D. W., & Krebs, J. R. (1986). Foraging theory (Vol. 6). Princeton university press.

51. Stephens, D. W., Brown, J. S., & Ydenberg, R. C. (Eds.). (2008). *Foraging: Behaviour and ecology*. University of Chicago Press.

52. Stoddart, D. M. (2012). *The ecology of vertebrate olfaction*. Springer Science & Business Media.

53. Vanak, A. T., Thaker, M., & Gompper, M. E. (2009). Experimental examination of behavioural interactions between free-ranging wild and domestic canids. Behavioural Ecology and Sociobiology, 64, 279-287.

54. Ward C, Smuts B (2006) Quantity-based judgments in the domestic dog (Canis lupus familiaris). Anim Cogn 10:71–80

55. Weissburg, M., Smee, D. L., & Ferner, M. C. (2014). The sensory ecology of nonconsumptive predator effects. The American Naturalist, 184(2), 141-157.

56. Woodford, R. (2012). Feed your best friend better: Easy, nutritious meals and treats for dogs. Andrews McMeel Publishing.

57. Zentall, T. R., Steirn, J. N., & Jackson-Smith, P. (1990). Memory strategies in pigeons' performance of a radial-arm-maze analog task. *Journal of Experimental Psychology: Animal Behaviour Processes*, *16*(4), 358.

# Supplementary information

**Supplementary Table 1**. Behavioural differences across different experimental phases.

| Pair | P-Value | Test | Adjusted P-Value | Behaviour |
|---|---|---|---|---|
| A_vs_B | 6.65e-02 | Chi-Square | 1.996e-01 | ML |
| A_vs_C | 6.69e-03 | Chi-Square | 2.008e-02 | ML |
| B_vs_C | 7.89e-06 | Chi-Square | 2.369e-05 | ML |
| A_vs_B | 3.65e-03 | Chi-Square | 1.095e-02 | ET |
| A_vs_C | 4.48e-26 | Fisher | 1.344e-25 | ET |
| B_vs_C | 1.10e-11 | Fisher | 3.306e-11 | ET |
| A_vs_B | 3.38e-02 | Chi-Square | 1.016e-01 | FG |
| A_vs_C | 4.66e-01 | Chi-Square | 1.000e+00 | FG |
| B_vs_C | 5.34e-03 | Chi-Square | 1.603e-02 | FG |
| A_vs_B | 2.75e-03 | Chi-Square | 8.265e-03 | SH |
| A_vs_C | 1.00e+00 | Chi-Square | 1.000e+00 | SH |
| B_vs_C | 1.31e-06 | Chi-Square | 3.955e-06 | SH |
| A_vs_B | 6.50e-05 | Chi-Square | 1.951e-04 | PG |
| A_vs_C | 3.04e-06 | Chi-Square | 9.149e-06 | PG |
| B_vs_C | 4.47e-11 | Chi-Square | 1.343e-10 | PG |

| A_vs_B | 2.04e-01 | Chi-Square | 6.121e-01 | RB |
| A_vs_C | 1.25e-02 | Chi-Square | 3.775e-02 | RB |
| B_vs_C | 5.04e-04 | Chi-Square | 1.513e-03 | RB |
| A_vs_B | 3.98e-02 | Chi-Square | 1.195e-01 | CF |
| A_vs_C | 1.66e-02 | Chi-Square | 5.008e-02 | CF |
| B_vs_C | 6.13e-01 | Chi-Square | 1.000e+00 | CF |

**Supplementary Table 2**: Logistic Regression Results for Strategic Behaviour Time Investment

| Strategic Behaviour | Condition | Estimate | p-value | Significance |
|---|---|---|---|---|
| SN | B | 0.156 | 0.435 | |
| SN | C | 1.265 | < 0.001 | *** |
| ML | B | -0.142 | 0.305 | |
| ML | C | -0.006 | 0.973 | |
| ET | B | -0.938 | < 0.001 | *** |
| ET | C | -2.026 | 0.003 | ** |
| FG | B | -0.970 | < 0.001 | *** |
| FG | C | -1.024 | 0.005 | ** |
| SH | B | -0.336 | 0.138 | |
| SH | C | -0.889 | 0.003 | ** |
| PG | B | -0.671 | 0.004 | ** |
| PG | C | -1.270 | < 0.001 | *** |
| LI | B | -0.312 | 0.628 | |
| LI | C | 0.277 | 0.671 | |
| RB | B | -0.178 | 0.499 | |
| RB | C | -0.617 | 0.045 | * |
| CF | B | -1.053 | < 0.001 | *** |

| | | | | | |
|---|---|---|---|---|---|
| CF | C | -0.947 | < 0.001 | *** |
| UBNB | B | -0.615 | 0.312 | |
| UBNB | C | -0.456 | 0.607 | |
| NB | B | 0.391 | 0.585 | |
| NB | C | 0.449 | 0.546 | |

**Note:**

- *p < 0.05 (\*), p < 0.01 (\*\*), p < 0.001 (\*\*\*).*

**Supplementary Table 3- Inter-rater reliability for event and time investment data**

| Data type | Metric | Value | 95% CI | Test statistic | *p*-value |
|---|---|---|---|---|---|
| Event data (categorical) | Cohen's Kappa | 0.919 | – | $z = 36.9$ | < 0.001 |
| Time data (continuous) | ICC(A,2) | 0.954 | 0.939–0.965 | $F(231,182) = 22.5$ | < 0.001 |

**Supplementary Table 4- Behavioural Transition Probability Matrix**

| | CF | CW | DG | E | FG | L | ML | NB | PG | RB | S | SH | SHNF | UBNB |
|---|---|---|---|---|---|---|---|---|---|---|---|---|---|---|
| **CF** | 0.000 | 0.000 | 0.214 | 0.036 | 0.036 | 0.000 | 0.000 | 0.000 | 0.464 | 0.000 | 0.000 | 0.250 | 0.000 | 0.000 |
| **CW** | 0.000 | 0.000 | 0.000 | 0.500 | 0.000 | 0.000 | 0.000 | 0.000 | 0.000 | 0.500 | 0.000 | 0.000 | 0.000 | 0.000 |
| **DG** | 0.200 | 0.000 | 0.000 | 0.100 | 0.000 | 0.000 | 0.400 | 0.000 | 0.000 | 0.000 | 0.100 | 0.200 | 0.000 | 0.000 |
| **E** | 0.000 | 0.000 | 0.000 | 0.000 | 0.000 | 0.000 | 0.000 | 0.000 | 0.000 | 0.000 | 0.000 | 0.000 | 0.000 | 0.000 |
| **FG** | 0.062 | 0.000 | 0.000 | 0.062 | 0.188 | 0.000 | 0.062 | 0.062 | 0.188 | 0.125 | 0.062 | 0.125 | 0.000 | 0.062 |
| **L** | 0.375 | 0.000 | 0.000 | 0.250 | 0.125 | 0.000 | 0.000 | 0.000 | 0.000 | 0.000 | 0.250 | 0.000 | 0.000 | 0.000 |
| **ML** | 0.268 | 0.000 | 0.000 | 0.073 | 0.122 | 0.000 | 0.024 | 0.024 | 0.049 | 0.146 | 0.098 | 0.146 | 0.024 | 0.024 |
| **NB** | 1.000 | 0.000 | 0.000 | 0.000 | 0.000 | 0.000 | 0.000 | 0.000 | 0.000 | 0.000 | 0.000 | 0.000 | 0.000 | 0.000 |
| **PG** | 0.049 | 0.000 | 0.000 | 0.049 | 0.000 | 0.000 | 0.268 | 0.000 | 0.049 | 0.244 | 0.122 | 0.220 | 0.000 | 0.000 |
| **RB** | 0.095 | 0.048 | 0.000 | 0.238 | 0.095 | 0.000 | 0.095 | 0.000 | 0.190 | 0.000 | 0.000 | 0.238 | 0.000 | 0.000 |
| **S** | 0.045 | 0.000 | 0.000 | 0.023 | 0.068 | 0.182 | 0.500 | 0.000 | 0.068 | 0.023 | 0.045 | 0.045 | 0.000 | 0.000 |
| **SH** | 0.088 | 0.029 | 0.118 | 0.294 | 0.000 | 0.000 | 0.000 | 0.000 | 0.412 | 0.029 | 0.000 | 0.029 | 0.000 | 0.000 |
| **SHNF** | 0.000 | 0.000 | 0.000 | 0.000 | 0.000 | 0.000 | 0.000 | 0.000 | 0.000 | 0.000 | 1.000 | 0.000 | 0.000 | 0.000 |
| **UBNB** | 0.000 | 0.000 | 0.000 | 0.000 | 0.500 | 0.000 | 0.000 | 0.000 | 0.000 | 0.000 | 0.500 | 0.000 | 0.000 | 0.000 |

**Supplementary information 1.**

**Survival Analysis Results**

**Kaplan–Meier Survival Analysis by Condition**

**Summary of Kaplan–Meier Estimates:**

| Condition | n | Events | Mean Survival (s) ± SE | Median Survival (s) | 95% CI Median (s) |
|---|---|---|---|---|---|
| A | 54 | 53 | 19.94 ± 2.31 | 13.25 | 11.25 – 16.79 |
| B | 50 | 27 | 65.75 ± 4.05 | 85.62 | 52.01 – NA |
| C | 50 | 2 | 88.15 ± 1.29 | NA | NA – NA |

**Log-rank Test (Survival Differences Across Conditions):**

$\chi^2 = 160$, df = 2, $p < 2 \times 10^{-16}$

**Cox Proportional-Hazards Model (Condition + Gender)**

| Predictor | coef | HR (exp(coef)) | 95% CI HR | z | p | Significance |
|---|---|---|---|---|---|---|
| Condition B | -2.0451 | 0.129 | 0.076 – 0.220 | -7.540 | $4.68 \times 10^{-14}$ | *** |
| Condition C | -4.9125 | 0.007 | 0.002 – 0.031 | -6.673 | $2.50 \times 10^{-11}$ | *** |
| Gender F | 0.3510 | 1.420 | 0.906 – 2.227 | 1.530 | 0.126 | ns |

**Model Fit Indices:**

- Concordance = 0.845 (SE = 0.018)
- Likelihood ratio test: $\chi^2 = 144.7$, df = 3, $p < 2 \times 10^{-16}$
- Wald test: $\chi^2 = 85.66$, df = 3, $p < 2 \times 10^{-16}$
- Score (logrank) test: $\chi^2 = 164.4$, df = 3, $p < 2 \times 10^{-16}$

**Proportional Hazards Assumption (Schoenfeld Residuals):**

| Predictor | $\chi^2$ | df | p |
|---|---|---|---|
| Condition | 2.576 | 2 | 0.28 |
| Gender | 0.599 | 1 | 0.44 |
| GLOBAL | 3.074 | 3 | 0.38 |

($\alpha = 0.05$; *** = significant, ns = not significant; PH assumption not violated)

**Supplementary information 2.**

| Test | $\chi^2$ (Chi-squared) | df | p-value |
|---|---|---|---|
| Kruskal–Wallis (Latency × Phase) | 0.703 | 2 | 0.704 |

**Supplementary information 3.**

**Time Difference Between First Sniff and First Lick**

**Normality Test (Shapiro–Wilk):**

$W = 0.4095, p < 2.2 \times 10^{-16}$

**Kruskal–Wallis Test:**

$\chi^2 = 35.2643, df = 2, p < 0.001$

**Post-hoc Pairwise Comparisons (Bonferroni corrected):**

| Comparison | Z | p (uncorrected) | p (adjusted) | Significance |
|---|---|---|---|---|
| A – B | -4.8980 | $4.84 \times 10^{-7}$ | $1.45 \times 10^{-6}$ | *** |
| A – C | -4.8572 | $5.95 \times 10^{-7}$ | $1.79 \times 10^{-6}$ | *** |
| B – C | -1.1688 | 0.121 | 0.364 | ns |

($\alpha = 0.05$; *** = significant after correction, ns = not significant)

**Supplementary information 4.**

**Behavioural Network Properties Across Conditions**

**Significant Network Properties:**

| Property | Condition A | Condition B | Condition C |
|---|---|---|---|
| Density | 1.132 | 1.496 | 0.771 |
| Average Degree | 29.43 | 44.88 | 21.60 |
| Reciprocity | 0.182 | 0.427 | 0.507 |
| Centralization (Degree) | 3.296 | 3.667 | 2.923 |
| Number of Strongly Connected Components | 1 | 2 | 2 |
| Edge Connectivity | 1 | 0 | 0 |
| Vertex Connectivity | 1 | 0 | 0 |

**Network Clustering Z-Test Across Conditions**

**Z-Test for Significance of Observed Clustering Coefficient vs. Random Networks**

| Condition | Observed Average Clustering | Z-Score | p-Value | Significance |
|---|---|---|---|---|
| A | 0.912 | 7.28 | <0.001 | *** |
| B | 1.310 | 16.86 | <0.001 | *** |
| C | 1.281 | 13.67 | <0.001 | *** |

($\alpha = 0.05$; *** = significant; indicates that the observed clustering is significantly higher than expected by chance compared to randomized networks)

**Supplementary figure 1:**

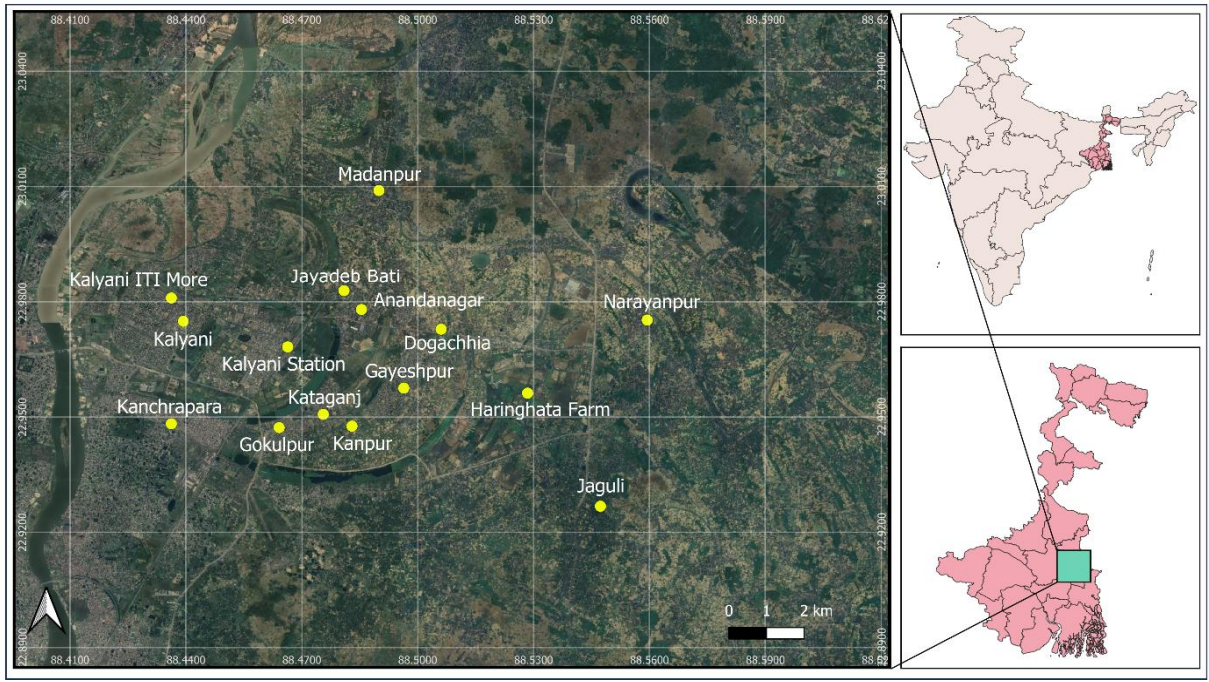

Fig. Study sites in Nadia district.

## Supplementary figure 2:

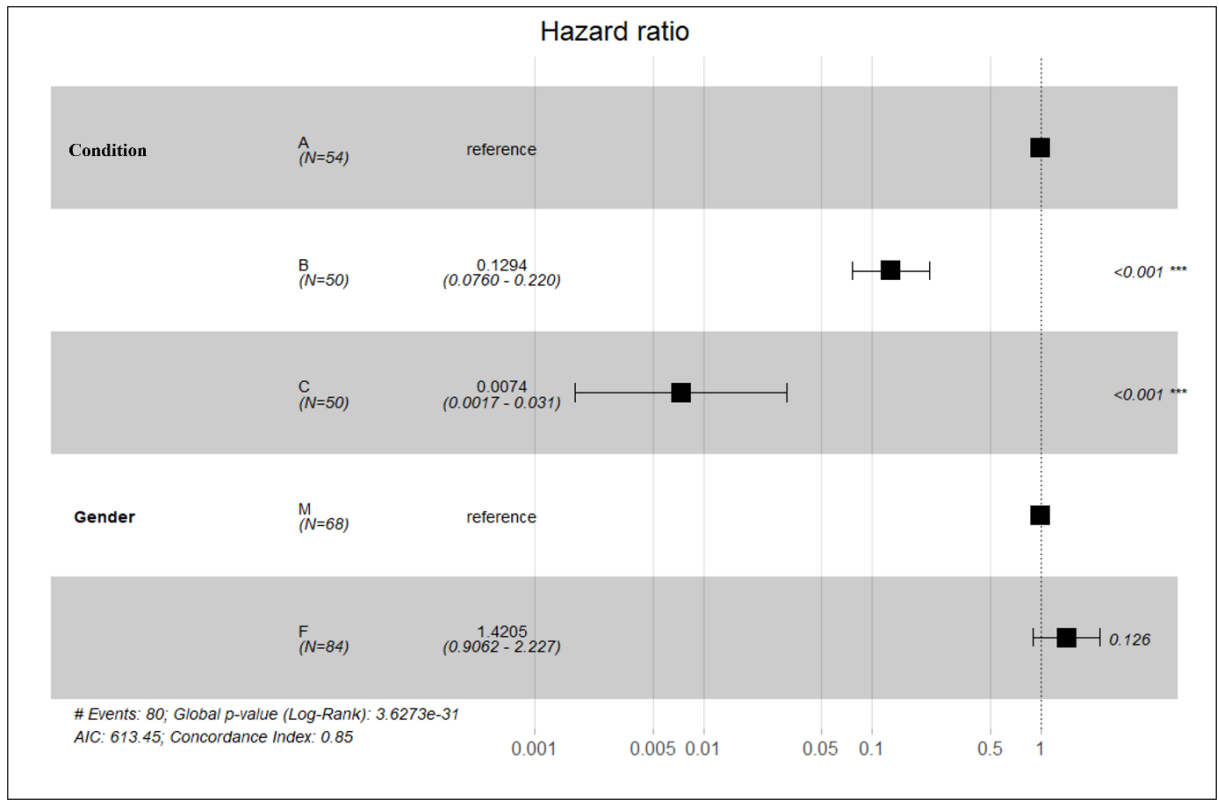

Figure: Hazard ratios from the Cox model showing differences in food acquisition times across conditions and genders. Conditions B and C had significantly lower hazard ratios than A (p < 0.001), indicating slower food consumption, while gender had no significant effect.

**Supplementary figure 3:**

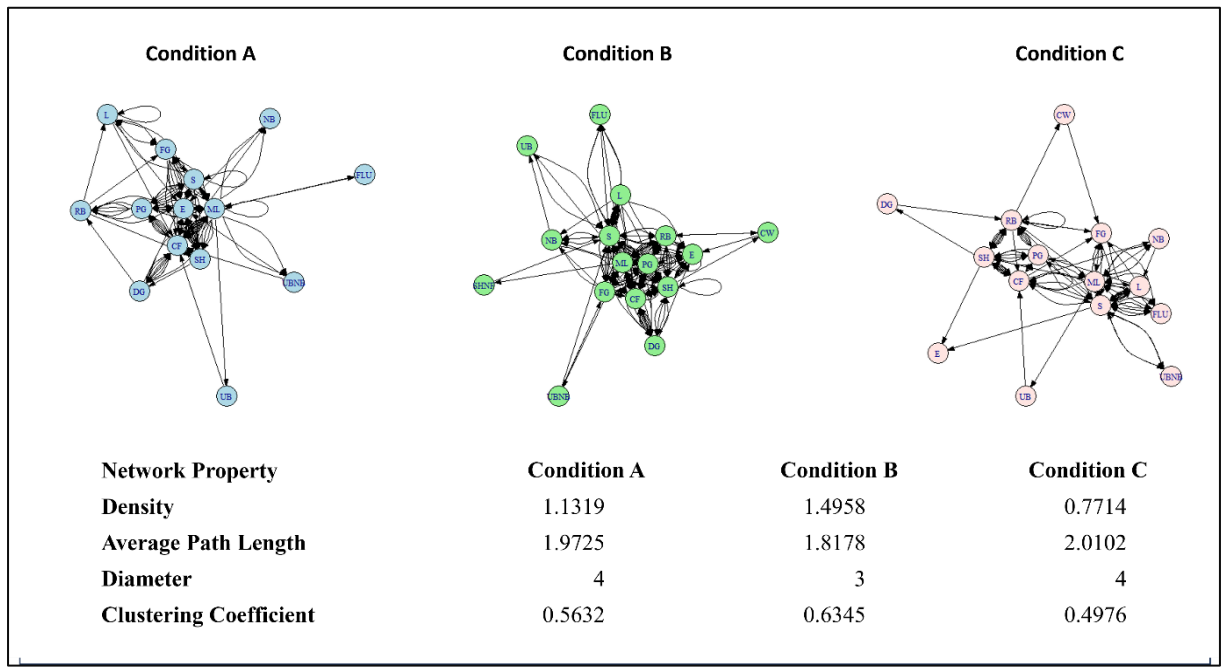

Fig. Network plots depicting behavioural interactions of free-ranging dogs across three experimental conditions.

**Supplementary figure 4:**

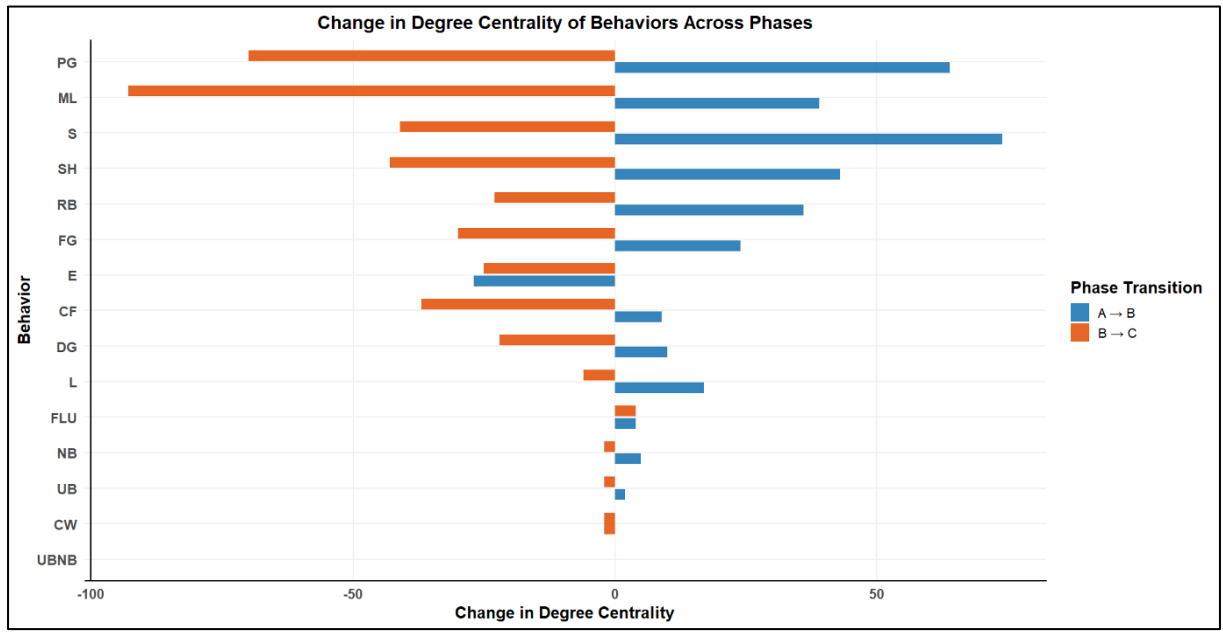

**Supplementary figure 5:**

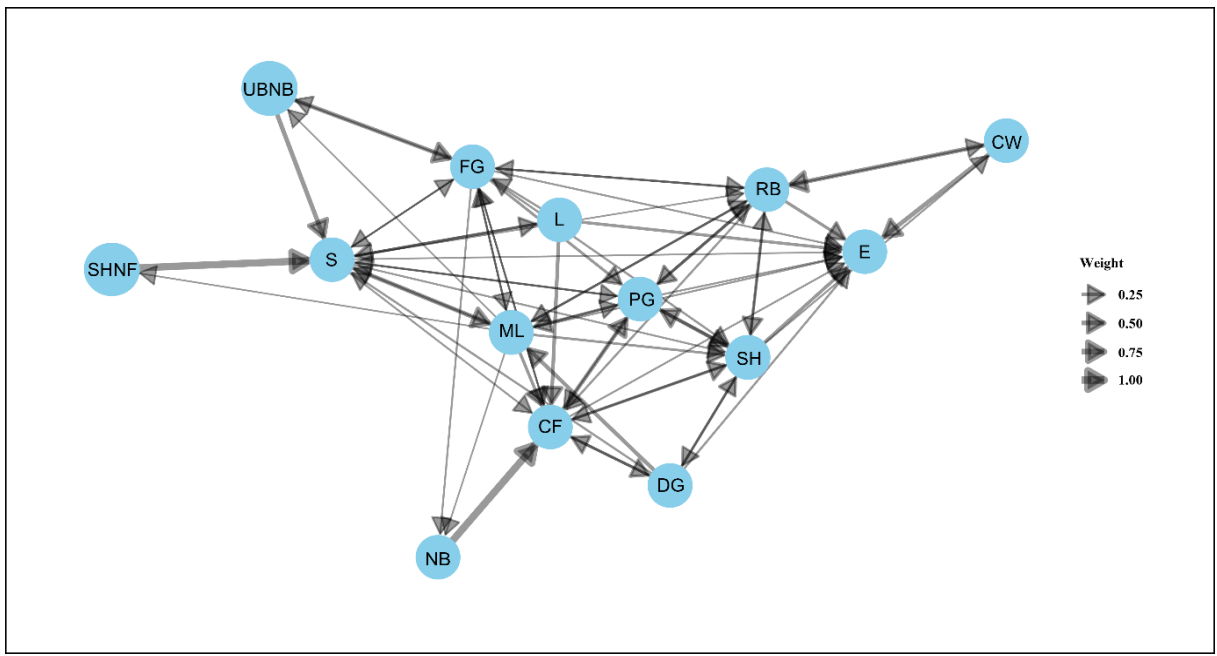

Fig. Behavioural transition network illustrating all observed transitions among behavioural states, where nodes represent behaviours and directed edges vary in thickness and colour intensity proportional to transition strength, highlighting stable high-probability behavioural routes while minimizing noise from rare transitions.

**Supplementary Video 1**: Experimental trial demonstrating the behavioural experiment

protocol, which was followed across Conditions A, B, and C.

https://drive.google.com/file/d/1hNYEYZn_J_-hTqMxteOMr9rx3DZm1H1A/view?usp=sharing.